\newif\ifconfver
\confvertrue        

\newif\ifplainver  
\plainvertrue

\newif\ifhide  
\hidetrue

\ifplainver
    \confverfalse   
\fi

\ifconfver
     \documentclass[10pt,twocolumn,twoside]{IEEEtran}
\else
    \ifplainver
        \documentclass[11pt]{article}
        \usepackage{fullpage}
    \else
        \documentclass[12pt,draftcls,onecolumn]{IEEEtran}
    \fi
\fi

\usepackage{etoolbox}%
\usepackage{xpatch}
\usepackage{blindtext}
\usepackage{tocloft}%
\usepackage{amsmath}

\newlength{\articlesectionshift}%
\let\LaTeXStandardSection\section
\let\LaTeXStandardTheSection\thesection
\let\LaTeXStandardTheSubSection\thesubsection
\let\LaTeXStandardTheSubSubSection\thesubsubsection
\let\LaTeXStandardTheParagraph\theparagraph

\makeatletter
\newcounter{titlecounter}

\xpretocmd{\maketitle}{\ifnumgreater{\value{titlecounter}}{1}}{\clearpage}{}{} 
\xpatchcmd{\maketitle}{\let\maketitle\relax\let\@maketitle\relax}{\refstepcounter{titlecounter}\begingroup
  \addtocontents{toc}{\begingroup\addtolength{\cftsecindent}{-\articlesectionshift}}%
  \addcontentsline{toc}{section}{\protect{\numberline{\thetitlecounter}{\@title~ \@author}}}%
  \addtocontents{toc}{\endgroup}
}{%
  \typeout{Patching was successful}
}{%
  \typeout{patching failed}
}%

\def\@IEEEdestroythesectionargument#1{\LaTeXStandardSection{#1}}%

\xapptocmd{\maketitle}{%
\renewcommand{\thesection}{\LaTeXStandardTheSection}%
\renewcommand{\thesubsection}{\LaTeXStandardTheSubSection}%
\renewcommand{\thesubsubsection}{\LaTeXStandardTheSubSubSection}%
\renewcommand{\theparagraph}{\LaTeXStandardTheParagraph}%
}{}{}%

\@addtoreset{section}{titlecounter}

\usepackage{calc,amsfonts,amssymb,amsmath,bm,url,color,theorem,graphicx,cite}
\usepackage{psfrag,float}
\usepackage{algorithm}
\usepackage{algorithmic}
\usepackage{soul}
\usepackage{enumerate}
\usepackage{bbm}
\usepackage{shortcuts_OPT}
\usepackage{multirow}
\usepackage{dsfont}

\usepackage{eqparbox}

\definecolor{orange}{RGB}{255,107,0}

\theorembodyfont{\rmfamily}

\usepackage{tikz}
\usetikzlibrary{arrows.meta}
\usetikzlibrary{decorations}
\usetikzlibrary{calc}
\usetikzlibrary{shapes.geometric}
\usetikzlibrary{external}

\begin{document}

\bibliographystyle{IEEEtran}

\newcommand{\papertitle}{
An Adversarial Model for Attack Vector Vulnerability Analysis on Power and Gas Delivery Operations
}

\newcommand{\paperabstract}{
    Power systems often rely on natural gas pipeline networks to supply fuel for gas-fired generation. Market inefficiencies and a lack of formal coordination between the wholesale power and gas delivery infrastructures may magnify the broader impact of a cyber-attack on a natural gas pipeline. In this study we present a model that can be used to quantify the impact of cyber-attacks on electricity and gas delivery operations. We model activation of cyber-attack vectors that attempt to gain access to pipeline gas compressor controls using a continuous-time Markov chain over a state space based on the gas operator Industrial Control System firewall zone partition. Our approach evaluates the operating states and decision-making in the networks using physically realistic and operationally representative models.
We summarize these models, the sequence of analyses used to quantify the impacts of a cyber-incident, and propose a Monte Carlo simulation approach to quantify the resulting effect on the reliability of the bulk power system by the increase in operational cost. The methodology is applied to a case study of interacting power, gas, and cyber test networks.\\

{\bf Keywords:} Continuous-time Markov chain, cyber-physical Systems, optimal power flow, Purdue model, transient gas flow

}


\ifplainver


    \title{\papertitle}

    \author{
    Ignacio Losada Carre\~no$^\dag$, Anna Scaglione$^\dag$, Anatoly Zlotnik$^\ddag$,\\ Deepjyoti Deka$^\ddag$ and Kaarthik Sundar$^\ddag$ \\ ~ \\
    $^\dag$Arizona State University, Tempe, AZ, USA, \\~ \\
    $^\ddag$Los Alamos National Laboratory, Los Alamos, NM, USA
    }

    \maketitle

    \begin{abstract}
    \paperabstract
    \end{abstract}

\ifconfver \else
    \ifplainver \else
        \newpage
\fi \fi

\section{Introduction}
\par Reliance on natural gas for electricity generation has increased significantly in many regions of the world \cite{EIA}. Low prevailing gas prices \cite{EIAGasPrice}, increasing penetration of renewable sources \cite{Renewables}, and the retirement of coal power plants \cite{Coal} have led natural gas to be the main source for energy in U.S. and European power grids. This creates an interdependence between power and gas delivery systems that compounds their reliability and security risks \cite{NERC_PowerGasCoordination}. It is becoming increasingly problematic for generation asset managers to procure natural gas during cold weather events \cite{PJMInterconnection2014} for base load and ancillary services, which may lead to load curtailments \cite{NEISOElectricityOutlook,NEISO-FuelSecurity}.

 \par Critically, the use of natural gas as a generation fuel source has altered the traditional status quo in which gas consumption is relatively steady and predictable throughout the day \cite{Zlotnik2015a,Zlotnik2017b}. This transition compels the need for analysis tools that incorporate transient pipeline flow modeling \cite{Zlotnik2016,Zlotnik2017,Zlotnik2017a}. A deficiency in formal coordination between power and gas delivery operations motivates the focus of our study on modeling the compounded impacts that may be caused by a malicious cyber attacker that seeks to maximize operational disruption across networks. 
 In some areas of the United States, a compressor station may be critical to supply natural gas to fuel up to 7,000 MW \cite{NEISO-FuelSecurity} of combined electricity generation capacity. An unplanned outage affecting these resources would disrupt grid operations, and thus it is crucial to understand the risk and consequence of a meditated attack on such an asset.
 \par In general, cyber-security threats to energy delivery systems appear in the form of unauthorized commands (e.g. opening a circuit breaker in the power system, a control valve in the natural gas network, or by changing relay settings), or of obfuscation of situational awareness (e.g. false data injection attacks that lead to sub-optimal operator decisions). In the former case, an attacker who is willing to disrupt operations may use a known or zero-day vulnerability to gain root access to a human machine interface (HMI). Open commands could be sent to a breaker and cause other relays to trip more lines if remedial actions are not taken in time. In the latter case, malicious agents may target communication links that are used to deliver data packets through the supervisory control and data acquisition (SCADA) network. Compromising a single communication link may affect control over multiple measurements, because measurement data is multiplexed at the remote terminal unit (RTU).

 \par Network security analysis involves the specification of attack graphs and trees. Attack graphs may occur in various forms, and attack vectors are challenging to model. There exists a substantial literature on analysis of attack graphs using Bayesian networks \cite{poolsappasit2011dynamic,zhang2015power,miehling2015optimal}. These studies use security metrics based on graph theory \cite{vukovic2012network} or vulnerability metrics \cite{boyer2007ideal,black2008cyber} such as the common vulnerability scoring system (CVSS) to study vectors of attack. However, these previous studies do not consider the importance of the evolving physical state and control protocols of the system, and do not examine the physical damage that could result in an energy delivery network. We argue that in a cyber-physical systems (CPS), cyber security is best understood in the context of models that simulate the underlying physical evolution and real-time operational control. Physical and economic metrics can then quantify the impacts of a given attack vector on an industrial control system (ICS). Our main contribution is a modeling approach that incorporates these considerations when evaluating cyber-attack impacts on infrastructure systems.

This study presents a stochastic adversarial model for quantifying vulnerability of interdependent gas-electricity delivery infrastructures to operational disruptions caused by cyber-attacks. Our approach is inspired by a previous study \cite{davis2015cyber}, which examined attack vectors for power systems as a Markov decision process (MDP) that represents attackers attempting to penetrate a corporate network through the internet, and which is used to emulate the behavior of a strategic agent. In that model, attackers may gain access to a power system control room from where actions (i.e. opening a circuit breaker) may be performed. Our study focuses on actions by non-strategic attackers to gain access and control of a gas pipeline network that provides fuel for generation that constitutes a significant portion of the generation mix of a dependent power grid. We examine how cyber-physical attacks on gas pipeline components can affect gas flow dynamics, subject to pipeline engineering and operating constraints, in order to quantify impacts on the ability of the pipeline to service gas-fired power plants and, in turn, the consequences on electric power systems security. Our approach is intended to enable methods to uncover vulnerabilities that arise from the interdependence between these energy systems. In particular, we seek to represent attack vectors that penetrate a gas cyber-network by acquiring credentials that allow the attacker to penetrate the cyber-physical infrastructure to attain direct control of operating points of pipeline gas compressors. The ability of a malicious cyber-attacker to access and control system components is represented using a finite-horizon continous-time Markov chain (CTMC) \cite{anderson2012continuous,yin2012continuous} that includes both the physical and cyber states. We propose that an energy management system (EMS) contingency analysis module could explore the likelihood of cyber-compromise events using a state-space approach, which could could be used offline in a N-k probabilistic outage model. Our formalism could be extended to include countermeasures, such as intrusion detection systems (IDS), honeypots, tamper-proof authentication, restrictive firewall rules, etc. We build a benchmark test case based on the Purdue network architecture model \cite{PurdueModel}. Finally, we attempt to capture a more realistic attacker behavior by incorporating the power and gas dynamics into the cyber modeling.

This manuscript is organized as follows. Section II presents the standard for hierarchical secure network architectures and describes interactions among cyber nodes and the physical system. We describe a cyber-gas-power workflow and the exchange of information among modules in an adversarial model for attack vectors on power and gas delivery operations. Sections III and IV, respectively, detail the transient gas and optimal power flow models used to evaluate operational impacts of attack vectors. Section V present the results of a computational study, and we conclude in Section VI.

\section{Cyber modeling using Markov Chains}\label{sec.ctmc}
\par
The cyber-topology of an ICS is represented by a set of firewall rules that limit the ability of third parties to control operations.
The firewalls and the rules that control access between zones define a graph in which nodes are the zones that are protected by firewalls and where each edge represents the possibility of devices from a one zone to penetrate the firewall that defends the next zone. We refer to this graph as the {\it cyber-topology}.
A cyber topology contains zones that can include the {\it enterprise network}, the {\it manufacturing zone} (e.g. the EMS), and the {\it area zone} (e.g. the SCADA network). The Purdue model \cite{PurdueModel} sets the current standard for firewalls that secure architectures in an ICS.
All IT systems are located in the enterprise zone, which in the case of energy delivery systems may include applications such as capacity planning, maintenance scheduling, e-mail, phone, and printing services, as well as VPN and corporate internet connections \cite{didier2011converged}. Because of security concerns, the Purdue model restricts communications between the enterprise and manufacturing zones to the so-called demilitarized zone (DMZ). In power and natural gas transmission systems, the manufacturing zone is responsible for controlling operations, and applications in this zone would include the plant historian, EMS, and file servers, which are core to the reliability and security of the power systems. The controls resulting from solving mathematical problems in the manufacturing zone (e.g. a unit commitment and/or economic dispatch in the power system) are then passed to the area zone where RTUs or programmable logic controllers (PLCs) send commands in real-time to actuators on the system. Information from metering devices, the status of switching devices, and alerts are displayed in workstations by human machine interfaces (HMIs). We propose that interactions between cyber-nodes and physical states of an energy system can be modelled using a continuous-time finite-state Markov process (i.e. chain), with states corresponding to penetration of the attacker into a specific zone of the ICS architecture. In this way, the CTMC model states and state transition matrix correspond to the cyber-topology of the ICS architecture.

\subsection{Continuous-time Markov chain model of attack vectors}
We express the Purdue model architecture as an undirected graph $\mathcal{G}_c=(\mathcal{V}_c,\mathcal{E}_c)$ where $\mathcal{V}_c$ denotes the set of cyber assets and $\mathcal{E}_c$ are the set of edges or communication links. That is, an edge connecting two nodes means that bi-directional traffic is allowed between the nodes. Cyber attackers are modeled as agents that follow a stochastic process $\{x(t) : t \geq 0\}$ within a CTMC \cite{anderson2012continuous,yin2012continuous}. A CTMC is represented as a tuple ($\mathcal{S},P(t),Q,\pi(t),\lambda$) defined on a continuous time interval $\mathcal{T}=[0,T]$ with an exponential holding time distribution $T_i \approx Exp(\lambda_i) \ \lambda_i\geq 0 \ \forall i \in \mathcal{S} $, where $\mathcal{S}=\{1,2,\dots,|\mathcal{S}|\}$ denotes the discrete state space. There exists a one-to-one mapping between every state $s \in \mathcal{S}$ and its corresponding cyber asset $i \in \mathcal{V}_c$. In addition, full control of the device is granted to the agent when a state from the Markov chain is realized. That is, the access to the area zone implies control over the substation actuators. Additionally, $\pi(t)=\big(p(x(t)=1),p(x(t)=2),\dots,p(x(t)=|\mathcal{S}|)\big)$ is the vector of probabilities of being at a given state at time $t$. The matrix $P(t) \in [0,1]^{|\mathcal{S}|\times|\mathcal{S}|}$ denotes the transition probability matrix and is defined for all times $t\in \mathcal{T}$. The set of communication links $\mathcal{E}_c$ provides the structure to $P(t)$, therefore the probability of transitioning between two states where no communications are allowed is 0. $Q \in \mathbb{R}^{|\mathcal{S}|\times|\mathcal{S}|}$ is the transition rate matrix which accounts for holding times. The transition rate matrix $Q$ can be obtained from the jump chain probability as
\begin{equation}
 Q_{i j}=\left\{\begin{array}{ll}{\lambda_{i} p_{i j}} & {\text { if } i \neq j}, \\ {-\lambda_{i}} & {\text { if } i=j,}\end{array}\right.
\end{equation}
where $p_{ij}$ is the embedded Markov jump probability between the $i$-th and $j$-th states. Note that $p_{ij}\neq[P(t)]_{ij}$, because the system evolves with time. Here $p_{ij}$ denotes the probability of an attacker to bypass a firewall rule. The state space contains a recurrent absorbing state that corresponds to the detection of the attacker by the operator, and the remaining states are transient. 
Therefore, there is a unique stationary point distribution. We determine $p_{ij}$ according to a security score that rates the likelihood of a given asset vulnerability to be exploited. Security scores can be found in ICS-CERT \cite{ICS-CERT} under advisories \cite{davis2015cyber}. The holding time parameters $\lambda_i$, $\forall i \in \mathcal{S}$, depend on the level of activity in the network. Also, calculating $P(t)$ for the continuous-time case is more challenging than in the discrete-time problem where $P(t)=P^t$ for $t\geq1$. The transition probability dynamics evolve according to
\begin{equation}\label{Kolmogorov-forward}
 \frac{d}{dt}P(t)=P(t)Q,
\end{equation}
with initial condition $P(0)=\mathbb{I}$ (i.e., no transitions are allowed at the initial state). The equation \eqref{Kolmogorov-forward} denotes Kolmogorov's forward differential equation \cite{Kolmogoroff1931}. The solution to \eqref{Kolmogorov-forward} is
\begin{equation}
 \begin{aligned} P(t)=e^{tQ} \equiv \sum_{n=0}^{\infty} \frac{(t Q)^{n}}{n !}. 
 \end{aligned}
\end{equation}

For convenience, it is assumed that initially the attacker resides on the internet, i.e., $\pi(0)=(1,0,\dots,0)$. Attacks may occur as a series of escalating vulnerability exploitations in the communication infrastructure, which may lead the attacker to the control room. Attackers may control the area zone (e.g. a substation RTU or PLC) and may alter controls sent to the field devices and change the setpoints. Software of various cyber assets such as RTUs or PLC devices are provided by vendors and may contain known vulnerabilities to an attacker \cite{ICS-CERT}. The attacker may also change the factory settings of a sensing device (i.e. a relay) provided by the vendor. This relay may send an open signal command to the circuit breaker due to bad modified settings. The consequences may include a redistribution of flows due to a change in the system topology, or an imbalance in the system if the generator setpoint is changed. Additionally, the HMI may be displaying false measurements, and may induce operators to believe that the system is in a different operating state for which suboptimal actions might be taken. The impacts of a cyber attack may compound the consequences of flaws in system design (e.g. resulting from inadequate transmission or capacity planning) experienced during extreme events.

The range of possible actions taken by a malicious agent with control access to an energy system (i.e. the discrete range of generator/compressor setpoints, the status of a circuit breaker/transformer, or the discrete interval of relay settings) are represented as states in the CTMC. In other words, the CTMC state space is characterized by the cyber and physical topology (i.e. the different hierarchical levels corresponding to the Purdue model, and substations) as well as set of actions an attacker may take to disrupt physical operations. An example of the CTMC graph is shown in Fig.~\ref{workflow}

 When the network gets attacked, i.e. when the Markov chain lands in a state that has the ability to control operations in real-time, a sequence of problems are solved to evaluate the operational impact of such attacks as presented in the following section. In this paper, we consider two ICS, namely the gas and electric grid interactions in the presence of a cyber-attack to the gas network.
\subsection{Modelling the interactions between power, cyber and gas}
Electricity and natural gas delivery operations rely on different practices to ensure reliability, resiliency, and security. Operator decisions can maintain normal functions of the system even given congestion and unexpected conditions. Therefore, the modelling of physics and real-world practices is paramount in assessing the impact of a cyber event. In the electric power sector, operations typically rely on the solution of two core optimization problems: 1) security-constrained unit commitment (SCUC), which is used for day-ahead scheduling of generators and reserve procurement; and 2) a DC optimal power flow (DCOPF), which is used to re-dispatch fast reacting units (e.g. gas turbines) in near real-time to compensate for imbalances arising from imperfect day-ahead forecasts. While the unit commitment problem in power systems is a well established framework (see e.g. \cite{tinney1967power,kerr1966unit,cohen1987optimization,wood2013power} and ``security-constrained'' variations of the problem \cite{shaw1995direct,monticelli1987security}), gas pipeline operations usually depend on labor-intensive heuristic approaches. Physical gas flow control is often reactive and may be sub-optimal. The use of optimization for pipeline operational scheduling under transient conditions is an active area of research \cite{Zlotnik2015a,Zlotnik2016,rios2015optimization,Zlotnik2017}.

\begin{figure*}
 \centering
 \includegraphics[width=\linewidth]{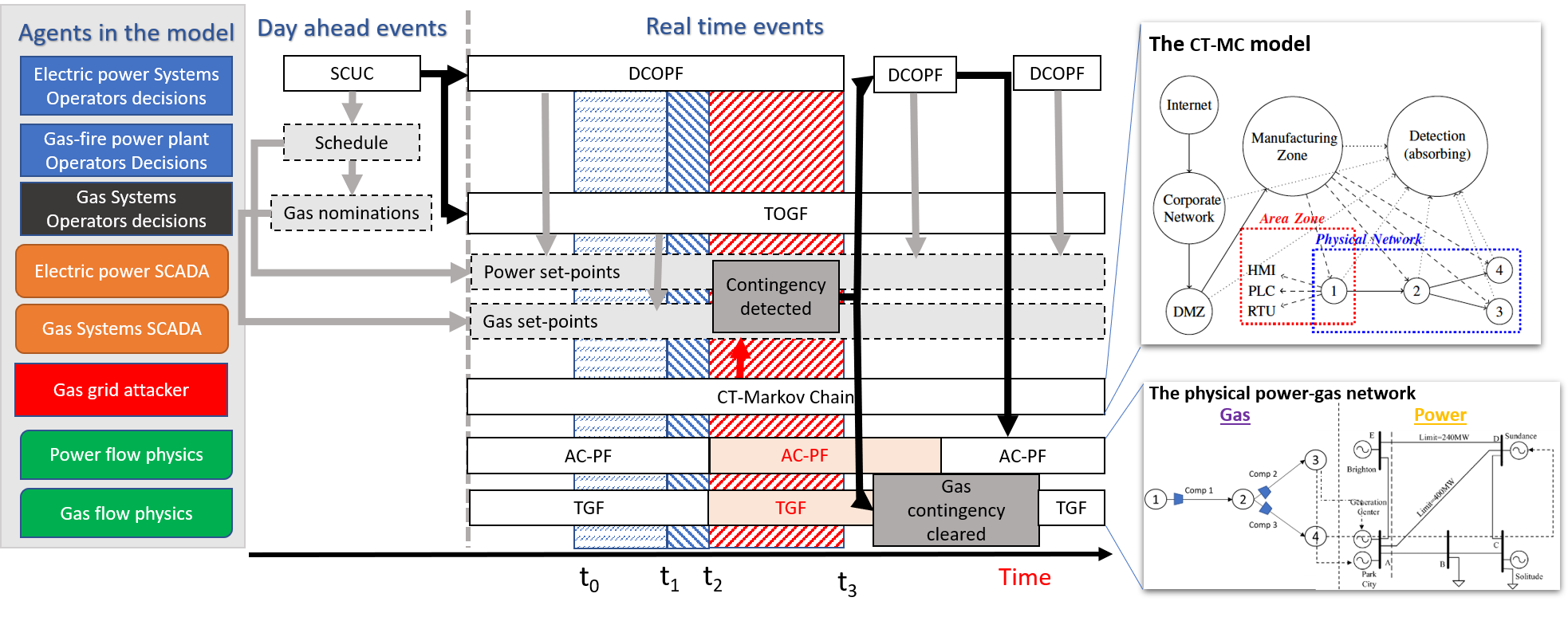}
 \caption{Summary of power, cyber, and gas network modelling. On the left side, we show the agents modelled in the problem that include operators, SCADA communications, the attacker, and the governing physics for the power grid and gas pipeline network. In the center of the figure, each block denotes our model for the decisions or action taken over time in sequence aligned vertically with the various agents that are operating in the coupled cyber-physical system. The times when the attacker penetrates the system, enters the control area zone, causes a contingency, and is detected are denoted by $t_0$, $t_1$, $t_2$, and $t_3$, respectively. On the right, we show an example of a Markov chain where the state space includes each of the areas in the Purdue model. Each area zone is coupled with a physical gas node of a 4-node pipeline network test case (i.e. a substation) where HMIs, PLCs and RTUs are located. Dashed edges connect cyber and physical nodes. Dotted edges denote the transitions to the detection state. Below the CTMC, a sample case of interdependent power-gas networks is shown. The blocks that are grayed-out are not explicitly coded in our numerical model. }
 \vspace{-3ex}
 \label{workflow}
\end{figure*}

\par The workflow of the mathematical problems we solve to study the impacts of cyber attacks is presented in Fig.~\ref{workflow}. The SCUC provides the schedules and natural gas nominations for power plants in the day-ahead markets, and the DCOPF in real-time re-dispatches generation as needed. For simplicity, we model synchronous 24-hour market periods, although this does not hold in practice in many regions. Because pipeline operations rely entirely on the judgement of experienced operators rather than on optimization-based decision support, we use transient optimal gas flow (TOGF) as a {\it best case} proxy for the actions of gas control operators to determine compressor operating setpoints. To evaluate the operational state of the pipeline system after a contingency, we use a transient gas flow (TGF) initial boundary value problem (IBVP) simulation. In the cyber domain, we model attack vectors through one CTMC, as is illustrated in Fig. \eqref{workflow}, that corresponds to the cyber-topology of the natural gas network. In this study, we only consider the case of attacks that attempt to access and control compressor stations, but the model is easily extensible to attacks on other system components as well as on the power system. When an attacker causes a compressor contingency, as observed in a transition of the CTMC to an appropriate state, we suppose that the boost ratio of the compressor drops to unity, so that downstream pipeline pressure may drop below minimum levels. If the pressure is sufficiently reduced, the pipeline operator will detect the event and issue a warning to the power system operator. When pressure is not restored to adequate values, gas supplies to power plants will be curtailed. This will require shutdown of gas-fired generators, which may be providing operational reserves to the power grid. In response, the power system will re-solve a DCOPF to dispatch spinning reserves and clear the contingency. However, power system reliability practices that require the N-1 security criteria could be insufficient if multiple gas-fired generators rely on natural gas delivered by a pipeline that depends on a single compressor, and thus, may not maintain sufficient line pressure \cite{sundar2016unit}. Also, operational generation reserves may not be available when the contingency occurs if they are procured from gas-fired power plants. Ultimately, such a chain of events will result in power load curtailment. In the following sections, we describe the conceptual and mathematical approaches, as well as numerical methods, that we use to simulate the decisions that power grid and gas pipeline operators would take for active management of interacting systems under attack.

\section{Modeling of gas system operations}\label{sec.gas-operations}
Historically, consumption of natural gas has been slowly varying and highly correlated to the weather \cite{edwardson2016next}. Natural gas pipeline managers have used simple capacity-based models to evaluate the ability of natural gas pipelines to support transport contracts in the day-ahead market \cite{Zlotnik2017}. The increase of natural gas usage for electricity generation has created significant intra-day fluctuations in gas consumption. Gas pipeline operators may rely on operating processes that assume steady ratable use of natural gas \cite{Zlotnik2015a,wong1968optimization,rothfarb1970optimal,luongo1991optimizing}, which do not accurately capture large fuel offtake ramps caused by the use of gas-fired generators to balance the variability and uncertainty in net-load \cite{pambour2018value}. This compels development and use of models that capture transient gas flow phenomena.

\subsection{Modelling transient gas dynamics in a network}
Recent studies have proposed new transient simulation and optimization tools that model intra-day transient gas dynamics in large-scale pipelines, their effect on pipeline capacity, and the consequences for dependent electric power systems \cite{Zlotnik2017a}. We consider the case of a gas network where gas can be injected, compressed or withdrawn. An example of a single pipeline with a compressor station is shown in Fig. \ref{pipe}.
\begin{figure}
 \centering
 \includegraphics[width=.8\linewidth]{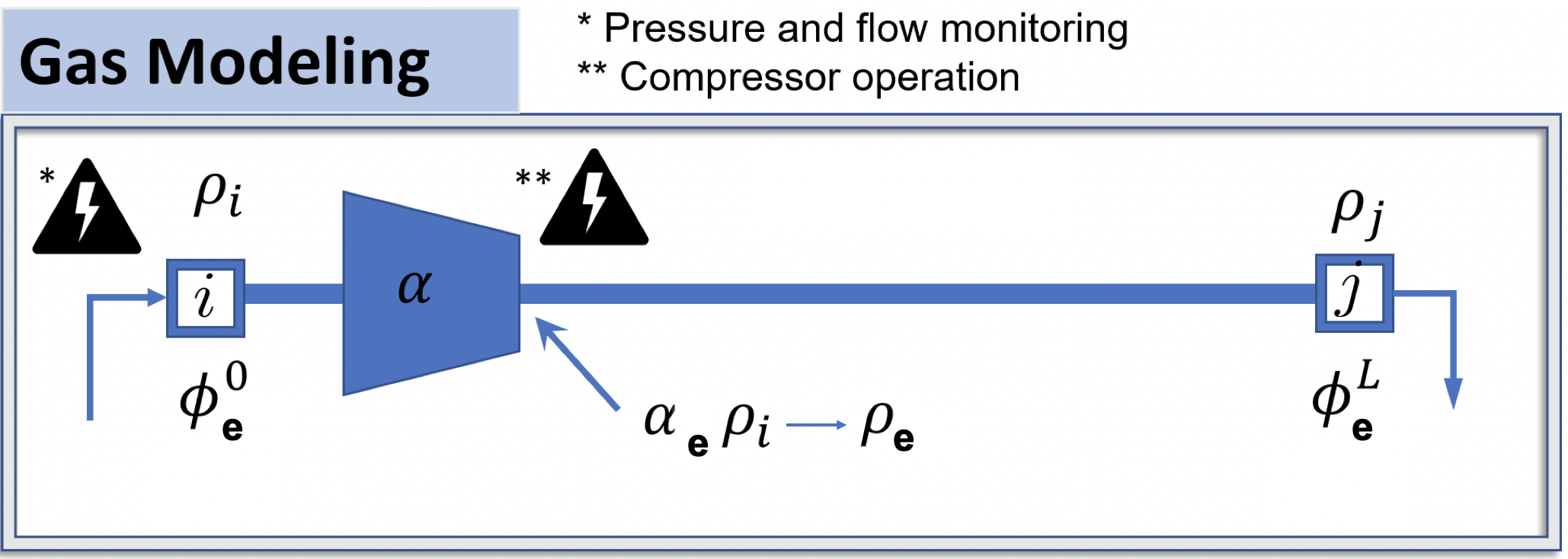}
 \caption{A pipeline with a compressor station}
 \label{pipe}
\end{figure}
A natural gas pipeline network can be represented as a directed graph $\mathcal{G}_g=(\mathcal{V}_g,\mathcal{E}_g)$ where $\mathcal{V}_g$ represents the set of nodes $i \in \mathcal{V}_g$ where gas can be withdrawn or injected. $\mathcal{E}_g$ represents the set of gas pipelines $e=(i,j)\in \mathcal{E}_g$ that connect two nodes $i,j$ where gas can be compressed with a ratio of outlet to inlet pressures $\alpha_{e}>1$, or transported (i.e., $\alpha_{e}=1$). Also, we denote $\mathcal{F}_i=\{(i,j) \mid (i,j)\in \mathcal{E}_g \ \text{and} \ i\in\mathcal{V}_g \ \forall j \} $, $\mathcal{T}_i=\{(j,i) \mid (j,i)\in \mathcal{E}_g \ \text{and} \ i\in\mathcal{V}_g \ \forall i \} $ as the set of edges connected to the $i$-th ``from'' or ``to'' node, respectively. We describe the evolution of compressible gas along a pipeline using a simplification of the 1-D Euler equations. We apply a non-dimensional transformation to the Euler equations in a single pipe following prior work \cite{Zlotnik2015a}, to yield
\begin{equation}\label{consmass}
 \underbrace{\partial_{t}\rho_{e}(t,x_{e})}_{\mbox{\tiny{Accumulation of mass}}}+\underbrace{\partial_{x}\phi_{e}(t,x_{e})}_{\mbox{\tiny{Net mass flux}}}=0,
\end{equation}
\begin{equation}\label{consmom}
 \underbrace{\partial_{t}\phi_{e}(t,x_{e})}_{\mbox{\tiny{Inertial forces}}}+\underbrace{\partial_{x}\rho_{e}(t,x_{e})}_{\mbox{\tiny{Pressure forces}}}=\underbrace{-\frac{\lambda_{e}\ell}{2D_{e}}\frac{\phi_{e}(t,x_{e}) |\phi_{e}(t,x_{e})|}{\rho_e(t,x_e)}}_{\mbox{\tiny{Friction forces}}},
\end{equation}
where \eqref{consmass} represents the continuity equation (conservation of mass) and \eqref{consmom} is the momentum conservation equation. The right hand side of \eqref{consmom} represents the Darcy-Weisbach equation where $\lambda$ is the Darcy friction factor. $\rho_e$ and $\phi_e$ denote the gas density and flux in the direction of motion ($x$), respectively. These variables are defined for the domain $[0,L_e]\times[0,T]$ being $L_e$ and $T$ the length of the pipe and time horizon. $D$ is the diameter of the pipe and $c$ denotes the speed of the sound defined by the ideal gas law as $p=c^{2}\rho$ where $p$ is the pressure. It should be noted that \eqref{consmass} and \eqref{consmom} assume slow dynamics ($v\leq15\ m/s$), neglect gravitational forces (horizontal pipes), and assume no temperature gradients (long pipes and high pressures). Density compatibility conditions related to pipe pressures are given by

\begin{equation}
 \rho_{e}(t,0)=\alpha_{e}(t)\rho_{i}(t) \quad \forall i: e \in \mathcal{F}_{i},
\end{equation}
\begin{equation}
 \rho_{e}(t,L_{e})=\rho_{i}(t) \quad \forall i: e \in \mathcal{T}_{i}.
\end{equation}
Constraints on nodal and compressor outlet densities are
\begin{equation} \label{denslim}
 \underline{\rho}_i\leq \rho_i(t) \leq \overline{\rho}_i \quad \forall i \in \mathcal{V}_g,
\end{equation}
\begin{equation} \label{denscomplim}
 \alpha_e(t)\rho_i(t)\leq \overline{\rho}_i \quad \forall i: e \in \mathcal{F}_{i},
\end{equation}
where $\underline{\rho}_i,\overline{\rho}_i$ denote the lower and upper density limits, respectively. Similarly, we impose limits on compressor ratios,
\begin{equation}\label{complim}
 1\leq \alpha_e(t)\leq \overline{\alpha}_e \quad \forall e \in \mathcal{E}.
\end{equation}
where $\overline{\alpha}_e$ is the compressor boost ratio upper limit.
The gas flow balance equation at a node is defined as
\begin{equation}\label{massbalance}
 d_{i}^g(t)=\sum_{e\in \mathcal{T}_{i}}A_e\phi_{e}(t,L_{e})-\sum_{e\in \mathcal{F}_{i}}A_e\phi_{e}(t,0), \quad \forall i \in \mathcal{V}_g,
\end{equation}
where $d_i^g$ denotes the net gas injection at the i-th node and $A_e$ is the cross-sectional area of the pipe $e$. Similarly we impose limits on gas injections or withdrawals with the constraints
\begin{equation}\label{sdlim}
 \underline{d}^g_i\leq d_i^g(t)\leq \overline{d}^g_i \quad \forall i \in \mathcal{V}_g
\end{equation}
where $\underline{d}^g_i$ and $\overline{d}^g_i$ are the minimum and maximum withdrawals at the node. In addition, the energy consumed by the compressor stations is given as
\begin{equation}\label{compobj}
 J_G=\!\!\!\!\sum_{\forall e\in \mathcal{E}}\!\!A_e\int_{0}^{T}\!\frac{|\phi_{e}(t,0)|}{\eta_{e}}(\alpha^{(\gamma-1)/\gamma}_{e}(t)\!-\!1)dt,
\end{equation}
where $\eta_e$ is a scaling factor that includes compressor efficiency and $\gamma$ is the gas specific heat capacity ratio \cite{menon2005gas}. We use a reduced control system model where the spatial dimension is discretized using a lumped-element scheme \cite{Zlotnik2015a}. Letting $N=|\mathcal{V}_g|$ and $E= |\mathcal{E}_g|$, we define the vector of nodal densities $\rho=\left(\rho_{1}^{L}, \ldots, \rho_{N}^{L}\right)^{T}$, the vectors of mass flux at the beginning and end of the pipe $\phi_{0}=\left(\phi_{1}^{0}, \ldots, \phi_{E}^{0}\right)^{T}$, $\phi_{L}=\left(\phi_{1}^{L}, \ldots, \phi_{E}^{L}\right)^{T}$, and injections or withdrawals $d^g=\left(d^g_{1}, \ldots, d^g_{N}\right)^{T}$. In addition, let $\Phi=\frac{1}{2}(\phi_{0}+\phi_{L})$. We define the weighted incidence matrix $B \in \mathbb{R}^{N \times E}$ by
\begin{equation}
 B_{i j}=\left\{\begin{array}{l}{1 \quad \text { edge } j \text { enters node } i, \text { i.e., } j=i} \\ {-\alpha_{j} \text { edge } j \text { leaves node } i, \text { i.e., } j=i+1} \\ {0 \quad \text { else }}\end{array}\right.
\end{equation}
where $A=\text{sign}(B)$ is the incidence matrix. Similarly, we define $A_s$ and $B_s$ as the incidence matrix corresponding to the nodes with known pressure, and $B_d$ and $A_d$ denote the remaining rows. We define the diagonal matrices $\Lambda$, $K$ and $X$ with diagonal entries $\Lambda_{ee}=L_e$, $K_{ee}=\ell \lambda_e/D_e$ and $X_{ee}=\pi D_e^2/4$ defined $\forall e \in \mathcal{E}_g$. In matrix-vector notation, the system dynamics \eqref{consmass}, \eqref{consmom}, and \eqref{massbalance} now become
\begin{equation}\label{vmfconsmass}
 \left|A_{d}\right| X \Lambda\left|B_{d}^{T}\right| \dot{\rho}=4\left(A_{d} X \Phi-d^g\right)-\left|A_{d}\right| X \Lambda\left|B_{s}^{T}\right| \dot{s}
\end{equation}
\begin{equation}\label{vmfconsmom}
 \dot{\Phi}=-\Lambda^{-1}\left(B_{s}^{T} s+B_{d}^{T} \rho\right)-K g\left(\phi,\left|B_{s}^{T}\right| s+\left|B_{d}^{T}\right| \rho\right)
\end{equation}
where the function $g : \mathbb{R}^{M} \times \mathbb{R}_{+}^{M} \rightarrow \mathbb{R}^{M} $ is defined by $g_{j}(x, y)=x_{j}\left|x_{j}\right| / y_{j}$. It should be noted that \eqref{vmfconsmass} is a result of combining \eqref{consmass} and \eqref{massbalance}. Refer to \cite{Zlotnik2015a,Zlotnik2015} for the full derivation of \eqref{vmfconsmass} and \eqref{vmfconsmom}. We use the above state space model to formulate and solve the following fundamental gas transport problems.
\subsubsection{Transient optimal gas flow (TOGF)}
We solve and optimal transient control problem where compressor ratios are minimized subject to the transient gas flow dynamic constraints and operating limits as follows
\begin{equation}\label{gasoptproblem}
\begin{array}{ll}
\!\!\!\! \min _{\alpha_{j}, \forall j} & \text{compressor energy }J_G \ \eqref{compobj} \\
\!\!\!\! \text{s.t.} & \text{system dynamics } \eqref{vmfconsmass}-\eqref{vmfconsmom} \\
& \text{pressure limits } \eqref{denslim}-\eqref{denscomplim} \\
& \text{compressor limits } \eqref{complim} \\
& \text{supply and demand limits } \eqref{sdlim}
\end{array}
\end{equation}

\subsubsection{Transient gas flow (TGF)}
We solve the differential equations \eqref{vmfconsmass}-\eqref{vmfconsmom} as an implicit Differential Algebraic Equation (DAE). The DAEs are structured as follows
\begin{equation}
 0=F\bigg(\begin{bmatrix}
 \dot{\rho}\\
 \dot{\Phi}
 \end{bmatrix},\begin{bmatrix}
 \rho\\
 \Phi
 \end{bmatrix},\begin{bmatrix}
 \alpha\\
 d^g
 \end{bmatrix}\bigg)=F(\underbrace{\dot{x}}_{\mbox{\tiny{gradient}}},\underbrace{x}_{\mbox{\tiny{state}}},\underbrace{u}_{\mbox{\tiny{controls}}})
\end{equation}
where $\rho$ and $\Phi$ are state variables, $x \in \mathbb{R}^{N+E}$, and $\alpha, d^g$ are the control variables $u \in \mathbb{R}^{|\mathcal{V}_g^d|+|\mathcal{E}_g^c|}$.

\section{Modeling of power system operations}
A power system network is represented as a graph $\mathcal{G}_p=(\mathcal{V}_p,\mathcal{E}_p)$ where $\mathcal{V}_p$ is the set of electric buses and $\mathcal{E}_p$ is the set of branches. Also, let $N_p=|\mathcal{V}_p|$ and $E_p= |\mathcal{E}_p|$. We formulate a SCUC problem for day-ahead scheduling and a DC-OPF for generation dispatch in real-time.
\subsection{Security-Constrained Unit Commitment (SCUC)}
North American Energy Reliability Corporation (NERC) standards require power systems to function reliably given an outage of any power system asset \cite{NERCreliability}. The SCUC formulation presented below guarantees that the power system is secure against a generator outage contingency. We solve an optimization problem defined on a discrete time interval $\mathcal{T}=\{1,2,\dots,N\}$ that optimizes operational cost:
\begin{equation} \label{ucobj}
 \begin{aligned}
 J^{uc}_P=\mathbf{P_{gt}^T}\mathbf{c_{g}}+\mathbf{u_{gt}^T}\mathbf{c_{g}^{nl}}+\mathbf{v_{gt}^T}\mathbf{c_{g}^{su}}+\mathbf{w_{gt}^T}\mathbf{c_{g}^{sd}} + \mathbf{r_{gt}^T}\mathbf{c^{r}_{g}}, \quad \forall t\in \mathcal{T},
 \end{aligned}
\end{equation}
where $\mathbf{P_{gt}} \in \mathbb{R}^{N_p}$ is the vector of power injections, $\mathbf{u_{gt}} \in \{0,1\}^{N_p}$ and $\mathbf{v_{gt}},\mathbf{w_{gt}} \in [0,1]^{N_p}$ are the vector of commitment, start-up and shut-down binary variables, respectively. We define $\mathbf{r_{gt}} \in \mathbb{R}^{N_p}$ as the vector of reserve capacities. Similarly, $\mathbf{c_{g}},\mathbf{c_{g}^{nl}},\mathbf{c_{g}^{su}},\mathbf{c_{g}^{sd}},\mathbf{c_{g}^{r}} \in \mathbb{R}^{N_p}$ are the vector of generator, no-load, start-up, shut-down, and capacity reserve costs, respectively. We use the subscript $t$ to index the elements of $P_g$ as provided in $\mathcal{T}$. The power flow thermal limits are
\begin{equation}\label{PFlim}
 \mathbf{\underline{P}_{\ell}} \leq PTDF^{R} \left(\mathbf{P_{gt}}-\mathbf{d_t^p}\right) \leq \mathbf{\overline{P}_{\ell}}, \quad \forall t\in \mathcal{T},
\end{equation}
where $\mathbf{d_t^p} \in \mathbb{R}^{N_p}$ is the vector of power withdrawals and $\mathbf{\underline{P}_{\ell}}$ and $\mathbf{\overline{P}_{\ell}}$ are the minimum and maximum power line thermal limits, respectively. $PTDF^R \in \mathbb{R}^{E_p\times N_p}$ denotes the power transfer distribution factor with respect to the reference bus, and can be calculated as $PTDF=HB'^{-1}$ where $H\in \mathbb{R}^{E_p\times(N_p-1)}$ is the power flow Jacobian and $B' \in \mathbb{R}^{(N_p-1)\times(N_p-1)}$ denotes the susceptance matrix excluding the row and column that correspond to the reference bus. Constraints on generator limits are given by
\begin{equation} \label{genlim}
 \text{diag}(\mathbf{\underline{P}}_{g})\mathbf{u_{gt}}+\mathbf{r_{gt}}\leq \mathbf{P_{gt}}\leq \text{diag}(\mathbf{\overline{P}}_{g})\mathbf{u_{gt}}-\mathbf{r_{gt}}, \quad \forall t\in \mathcal{T},
\end{equation}
where $\mathbf{\underline{P}}_{g}$ and $\mathbf{\overline{P}}_{g}$ are the minimum and maximum generator limits, respectively. The operator $\text{diag}(.)$ returns a diagonal matrix with the elements of the vector argument. Generation must balance system load at all times, as enforced by
\begin{equation}\label{pbalance}
 \mathds{1}^{\mathrm{T}} \mathbf{P_{gt}}=\mathds{1}^{\mathrm{T}} \mathbf{d_t^p}, \quad \forall t\in \mathcal{T},
\end{equation}
and we impose ramp rate limits on the generation:
\begin{equation}\label{rampuplim}
 \mathbf{P_{gt}}-\mathbf{P_{g,t-1}}\leq \text{diag}(\mathbf{R_{g}^{hr}})\mathbf{u_{g,t-1}}+\text{diag}(\mathbf{R_{g}^{su}})\mathbf{v_{gt}}, \quad \forall t\in \mathcal{T},
\end{equation}
\begin{equation}\label{rampdownlim}
 \mathbf{P_{g,t-1}}-\mathbf{P_{gt}}\leq \text{diag}(\mathbf{R_{g}^{hr}})\mathbf{u_{gt}}+\text{diag}(\mathbf{R_{g}^{sd}})\mathbf{w_{gt}}, \quad \forall t\in \mathcal{T},
\end{equation}
where $\mathbf{R_{g}^{hr}},\mathbf{R_{g}^{su}},\mathbf{R_{g}^{sd}} \in \mathbb{R}^{N_p}$ are the hourly, start-up, and shut-down ramping rates. The commitment constraints are expressed using binary variables, and are given as
\begin{equation} \label{minuptime}
 \sum_{s=t-\underline{t}^u_{g}+1}^{t}\mathbf{v_{gs}}\leq \mathbf{u_{gt}}, \quad \forall t \in \{\underline{t}^u_{g},\dots,N\},
\end{equation}
\begin{equation} \label{mindowntime}
 \sum_{s=t-\underline{t}^d_{g}+1}^{t}\mathbf{w_{gs}}\leq 1-\mathbf{u_{gt}}, \quad \forall t \in \{\underline{t}^d_{g},\dots,N\},
\end{equation}
\begin{equation} \label{commitmentupdown}
 \mathbf{v_{gt}}-\mathbf{w_{gt}}=\mathbf{u_{gt}}-\mathbf{u_{g,t-1}}, \quad \forall t\in \mathcal{T},
\end{equation}
where $\underline{t}^u_{g}$ and $\underline{t}^d_{g}$ are the minimum up and down times of generator $g$. Allocation of reserve capacity is guaranteed by imposing the constraints
\begin{equation}\label{res7percent}
 \mathds{1}^{\mathrm{T}}\mathbf{r_{gt}}\geq \mathds{1}^{\mathrm{T}}(0.07\mathbf{d^p_{t}}), \quad \forall t\in \mathcal{T},
\end{equation}
\begin{equation}\label{resanygen}
 \mathds{1}^{\mathrm{T}}\mathbf{r_{gt}}\otimes\mathds{1}\geq \mathbf{P_{gt}}+\mathbf{r_{gt}}, \quad \forall t\in \mathcal{T},
\end{equation}
\begin{equation}\label{rescommitment}
 \mathbf{r_{gt}}\leq \text{diag}(\mathbf{R_{g}})\mathbf{u_{gt}}, \quad \forall t\in \mathcal{T}.
\end{equation}
In other words, the system must be able to allocate enough capacity to cover the outage of any single generator or 7\% of the load at any given time, whichever is larger. The SCUC optimization problem is
\begin{equation}\label{ucpoweroptproblem}
\begin{array}{ll}
\!\!\!\! \min _{\mathbf{P_{gt}}} & \text{operational cost }J^{uc}_P \ \eqref{ucobj} \\
\!\!\!\! \text{s.t.} & \text{power flow limits } \eqref{PFlim} \\
& \text{generator limits } \eqref{genlim} \\
& \text{power balance } \eqref{pbalance} \\
& \text{ramping limits } \eqref{rampuplim}-\eqref{rampdownlim} \\
& \text{commitment constraints } \eqref{minuptime}-\eqref{commitmentupdown}\\
& \text{reserve capacity constraints } \eqref{res7percent}-\eqref{rescommitment}

\end{array}
\end{equation}

\subsection{DC Optimal Power Flow (DC-OPF)}
To emulate economic re-dispatch in real-time electricity markets, we use a DC-OPF formulation $\forall t \in \mathcal{T}=\{1,2,\dots,N\}$ where the commitment variables and reserves are passed as parameters. The optimization problem is
\begin{equation}\label{ucpoweroptproblem}
\begin{array}{ll}
\!\!\!\! \min _{\mathbf{P_{gt}}} & \mathbf{P_{gt}^T}\mathbf{c_{g}} \\
\!\!\!\! \text{s.t.} & \text{power flow limits } \eqref{PFlim} \\
& \text{generator limits } \eqref{genlim} \\
& \text{power balance } \eqref{pbalance} \\
& \text{ramping limits } \eqref{rampuplim}-\eqref{rampdownlim}

\end{array}
\end{equation}


\subsection{Physical coupling between power and gas infrastructures}
We model the interconnections between the wholesale gas and electricity sectors created by gas-fired generators using quadratic heat curves of the form
\begin{equation}\label{eq.gas-coupling}
 d_i^g[t]=\beta(a\cdot P^2_{it}+b\cdot P_{it}+c) \quad \forall i \in \mathcal{V}_p \cap \mathcal{V}_g, \quad \forall t \in \mathcal{T},
\end{equation}
where $a$, $b$, and $c$ are heat rate coefficients and $\beta : \text{MMBTU/MWh} \rightarrow \text{kg/s}$ is a conversion factor.
\section{Simulation Case Study and Results}
We present a study to showcase the numerical results that can be obtained by concatenating the modules described in Fig.~\ref{workflow}. One realization of the workflow is a stochastic process where the nonlinear evolution in the state of the energy networks depends on the time interval during which the Markov chain is in a state that represents compromised gas compressor operations. In order to quantify the distribution of the stochastic process and the resulting effect on energy delivery reliability, we perform Monte Carlo simulation of the stochastic process and examine its statistics.
\begin{table}[]
\centering
\caption{Coupling of pipeline nodes to gas-fired generators.}
\begin{tabular}{ccc}
\textbf{Gas Node} & \textbf{Generator ID} & \textbf{Total Capacity (MW)} \\
6 & 4,20,21 & 255 \\
8 & 5,22,23 & 255 \\
12 & 11,24,25 & 255 \\
13 & 12,26,27 & 255 \\
18 & 13,28,29 & 255 \\
19 & 15,31 & 231 \\
24 & 36,37,38 & 228 \\
25 & 39,40 & 152
\end{tabular}
\label{tab.coupling}
\end{table}

\textbf{Test System:} The cyber-gas-power numerical analysis is done on the IEEE-118 test case \cite{christiepower} for the power system coupled with a 25-node gas test case from \cite{Zlotnik2016} with five compressor stations. We model the coupling of gas-fired generators in the gas network as shown in Table \ref{tab.coupling}. The generator heat rate coefficients are shown in Table \ref{tab.heatrates}. The natural gas cyber-network follows the Purdue model discussed in Sec. \ref{sec.ctmc}, resulting in the CTMC shown in Fig.~\ref{workflow}. We use one absorbing state to denote the attack detection state. Attacks occur as a series of firewall vulnerability exploitations that lead to compressor station contingencies and are modelled by the transition probabilities. We assume that compressor stations with higher impact in the a radial gas network to have higher vulnerability scores, thus, reflecting the fact that some areas will be more difficult to access.
\begin{table}[H]
\centering
\caption{Gas-fired generator heat rates \cite{lew2012impacts}}
\begin{tabular}{cccc}
 & \textbf{a} & \textbf{b} & \textbf{c} \\
\textbf{Combustion turbine} & 4.46 & -9.95 & 15.11 \\
\textbf{Steam turbine} & 1.7 & -3.87 & 12 \\
\textbf{Combined cycle} & 5.8 & -11.2 & 12.87
\end{tabular}
\label{tab.heatrates}
\end{table}

\textbf{Implementation:} We use Julia/JuMP v0.19.2 to solve the optimization problems. JuMP is used to call the non-linear solver Ipopt \cite{wachter2006implementation} to solve the non-convex, non-linear TOGF, and Gurobi is called for the SCUC and DCOPF which are mixed-integer (MIP) and linear (LP) programs, respectively. The physical flows in the natural gas network are modelled using the TGF IBVP presented in Section \ref{sec.gas-operations}. The DAEs are solved by using the solver IDA in Sundials. IDA uses a variable-order, variable-coefficient Backward Differentiation Formula (BDF) with a modified Newton iteration method and is called through the Julia package ``DifferentialEquations.jl''. The initial conditions are obtained by solving a steady-state optimization problem, i.e. we set $\dot{\rho}=0$ and $\dot{\Phi}=0$ in \eqref{gasoptproblem}, for which we use an absolute tolerance of $10^{-5}$ and a relative tolerance of $10^{-6}$. We obtain 10,000 samples of the Markov chain over a 24-hour operational day through the Monte Carlo simulation. The total processing time is less than 30 minutes using an Intel i7 4-core processor 2018 Macbook Pro.

\begin{figure}[]
 \centering
 \includegraphics[width=0.7\linewidth]{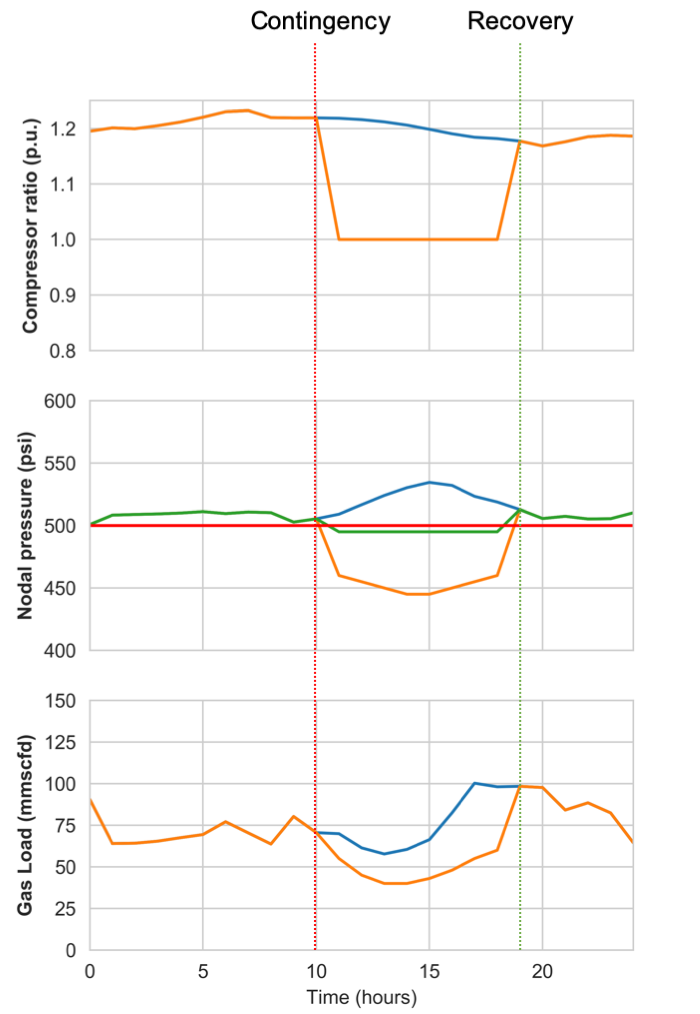}
 \caption{Sample of the cyber-gas-power simulation workflow shown for a node in the gas test network at node 6. Top: compressor ratio at station 3; center: nodal gas pressure; bottom: gas load at the node. TOGF solutions without contingency (blue), given a contingency (orange), and correction by generator curtailment (green).}
 \vspace{-3ex}
 \label{fig:workflow_sample}
\end{figure}

\textbf{Simulation workflow:} First, we compute a baseline SCUC and DCOPF solution for the power system, which results in a gas load at the node. Figure \ref{fig:workflow_sample} illustrates one instance of the physical system states at gas node 6 that services gas-fired generators 4, 20 and 21 with a combined capacity of 255 MW. Note that the nominal pressure at that node is just over 500 psi, which is the minimum pressure bound in the network. A TOGF is then used to compute the gas compressor profile and nodal gas pressure under normal operations (shown in blue in the figure). The CTMC is used to compute the timing of a compressor outage in the form of contingency and recovery times, between which an upstream gas compressor pressure boost ratio drops to $\alpha_e\equiv 1$. We then perform a TGF simulation to show the resulting effect on the nodal pressure (shown in orange in the figure). When the pressure drops below minimum levels, we suppose that the gas-fired generator is curtailed, which results in an acceptable nodal pressure (shown in green in the Figure). However, the production of the curtailed power plant must be replaced with more expensive resources, which increases the economic cost of production. We thus quantify the impacts of cyber-attacks by the resulting cost of reserves and by the percentage of electric load shed.

\begin{figure}[]
 \centering
 \includegraphics[width=0.6\linewidth]{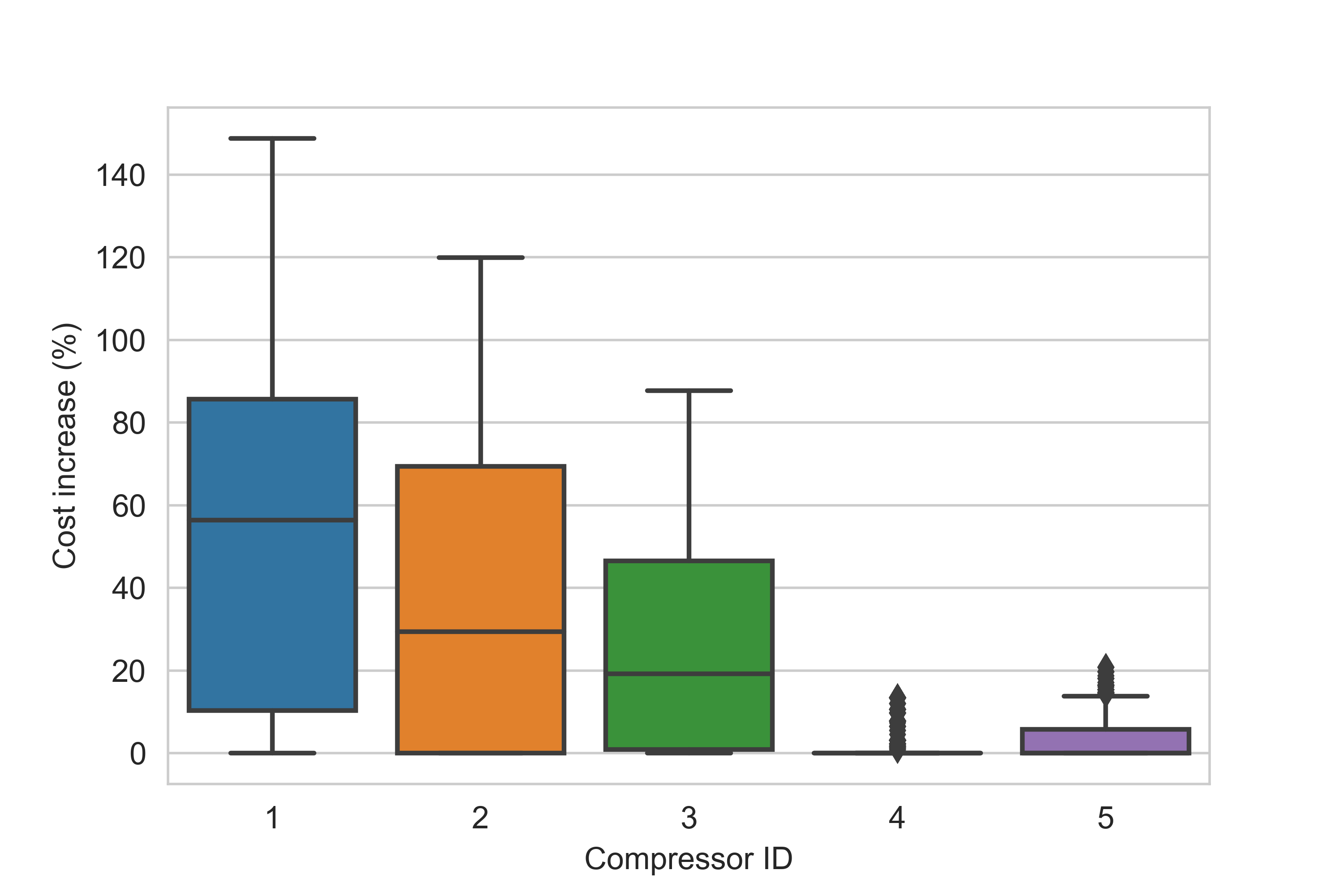}
 \caption{Compressor contingency boxplot. The y-axis shows the power operational cost increase as a percentage of the base case without contingencies}
 \label{boxplot}
\end{figure}
The range of outcomes that could result from a contingency in a natural gas compressor station is given in Fig.~\ref{boxplot}. The 25-node tree network heavily relies on compressor 1 to sustain a minimum pressure. A contingency in compressor 1 will result in natural gas fuel curtailments to all gas-fired generators. As a consequence, the power system operator will re-dispatch its most expensive generator to balance the system, deploy reserves and sometimes curtail power load with results in an increase in cost up to 150\%. In fact, such an outcome would be expected for actual pipelines, which have topological tree structures. Another scenario may involve an attacker targeting compressor 2, which provides natural gas to all generators connected to gas nodes 6 and 8. The power system is not compliant with events that cause multiple contingencies and thus such an event may compromise power system reliability. In Fig.~\ref{fig.gascorrelations}, we examine how curtailed gas loads correlate to the increase in operational cost for power systems, and also examine the power load curtailment with respect to the operational cost increase and curtailed gas loads.
\begin{figure}
    \centering
    \includegraphics{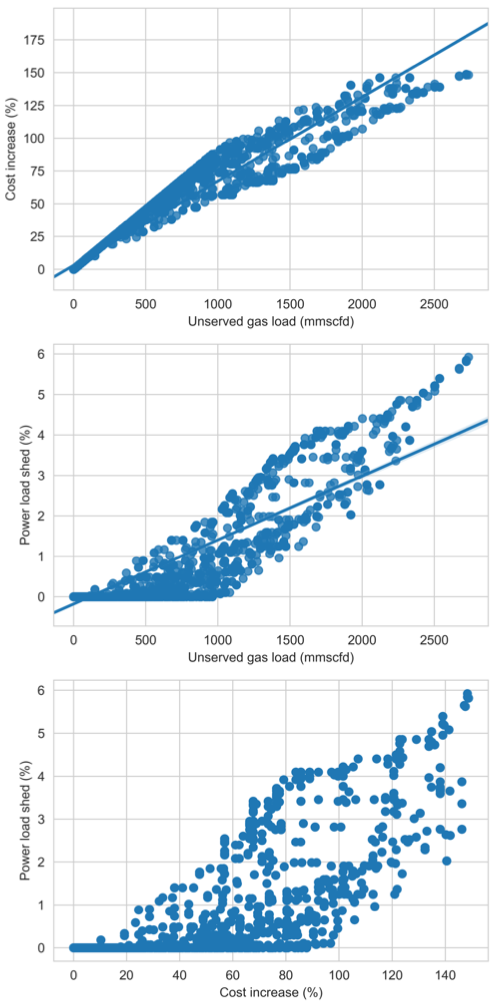}
    \caption{Top: Operational power cost increase vs. curtailed gas loads. Center: Power load shed vs. curtailed gas loads. Bottom: Power load shed vs. operational power cost increase}
    \label{fig.gascorrelations}
\end{figure}
We see a strong correlation between the operational cost and curtailed gas load ($R=0.9813$), and also observe that there exist cases where the increase in cost also results in the highest power load shedding.

\section{Conclusion}
This study provides a detailed simulation framework to rigorously analyze the impacts of malicious cyber-attacks on the physical and operating states of interdependent natural gas and electric power transmission infrastructures. We incorporate the power and natural gas dynamics into the cyber modeling to better capture the consequences of attacks on an energy system. The proposed modeling approach can be extended to accommodate any ICS with any firewall policy in place. Through a numerical study involving cyber attacks restricted to gas compressor contingencies in a pipeline, we show that such an attack can cause significant cost increase and disruption in the operations of the power grid. Our results suggest that N-1 security standards may not suffice to guarantee reliable power system operation. Additionally, coordinated power and gas delivery operations would increase the reliability of both sectors. Going forward, we will extend our attack construction to more potent adversaries that can target not just gas compressor operations, but other system components such as control valves, storage fields, or power system equipment. Our future work will also extend the modeling of cyber-attacker behavior by employing a Markov decision process that involves observed system states, rather than a single Markov chain only.
\section{Acknowledgements}
This study was carried out as part of the ``Dynamical Modeling, Estimation, and Optimal Control of Electrical Grid-Natural Gas Transmission Systems'' Project for the D.O.E. Office of Electricity Advanced Grid Research and Development program. Work conducted at Los Alamos National Laboratory was done under the auspices of the National Nuclear Security Administration of the U.S. Department of Energy under Contract No. 89233218CNA000001. It was also supported in part by the Director, Office of Electricity Delivery and Energy Reliability, Cybersecurity for Energy Delivery Systems program, of the U.S. Department
        of Energy, under contract DOE0000780. Any
        opinions, and findings expressed in this material are those of the authors and
        do not necessarily reflect those of the sponsors.

\bibliographystyle{IEEEtran}
\bibliography{Bibliography}

%


\end{document}